\documentclass[aps,prd,twocolumn,groupedaddress,nofootinbib]{revtex4-1}
\pdfoutput=1
\usepackage{graphicx}
\usepackage{dcolumn}
\usepackage{bm}
\usepackage{braket}
\usepackage{verbatim}
\usepackage{amsmath}

\begin{document}

\title{Properties of the (Un)Complexity of Subsystems}
\author{Henry Stoltenberg}
\email{henry-stoltenberg@oist.jp}
\affiliation{Okinawa Institute of Science and Technology,\\ 1919-1 Tancha
	, Onna-son, Okinawa 904-0495, Japan}

\begin{abstract}
I investigate some properties of proposed definitions for subsystem/mixed state complexity and uncomplexity. A very strong dependence arises on the density matrix's degeneracy which gives a large separation in the scaling of maximum subsystem complexity with number of qubits (linear compared to exponential). I also
investigate several cases where the uncomplexity of quantum states are superadditive and present some challenges and progress in showing that the relation holds in complete generality.

\end{abstract}

\maketitle

\section{Introduction and Background}
Quantum information has become increasingly important in understanding gravity, which has been realized in studying black hole information \cite{Hayden:2007cs, Almheiri:2012rt} and holography \cite{Ryu:2006bv, Faulkner:2013ica, Swingle:2014uza}. Going beyond the connections between entanglement and geometry probed by Von Neumann entropy, interest in a much more fine grained quantity, quantum complexity has grown.
Quantum state complexity has been conjectured to be dual to either the volume or action of a black hole's interior \cite{Stanford:2014jda, Brown:2015bva, Carmi:2017jqz}. Complexity potentially also plays an important role in resolving/understanding the formation or lack of firewalls for black holes \cite{Harlow:2013tf, Maldacena:2013xja}. 

An interesting parallel to entropy is that complexity seems to obey a second law \cite{Brown:2017jil} which suggests that formulating some analog of thermodynamics may be possible. The authors also point out that much like how free energy acts as a resource to preform useful work \cite{Gour:2013}, uncomplexity (remaining complexity) may be a resource for useful computations. 

In the context described above, the typical definitions of quantum complexity describe only unitary operations acting on pure states. Many systems of interest are subsystems of a larger pure state (e.g. evaporating black hole, half of the thermofield double state), and these subsystems will generally be mixed. In the holographic context, mixed subsystems would correspond to subregions. Subregion complexity has been calculated relying on the proposed holographic duality between complexity and volume \cite{Alishahiha:2015rta, Ben-Ami:2016qex}.
For quantum circuits, several possible definitions for mixed state/subsystem complexity have been recently proposed in \cite{Agon:2018zso}. What the authors define as purification complexity and basis complexity will be the focus of the following work. 
The goal of this paper is to better understand how the complexity of a system and its constituents are related. In particular I will investigate whether a superadditive property holds for uncomplexity.

\section{Complexity Definitions} 
The definitions of complexity that will be used in this paper will be built upon the typical definitions for pure state complexity. In AdS/CFT, the quantum complexity that we would want to describe would be for the state of the CFT. There have been several proposals of how to define pure state complexity in field theory \cite{Jefferson:2017sdb,Chapman:2017rqy}. Instead in this paper, I will restrict to the context of quantum circuits with qubits for simplicity.
I will always be describing a system of 2N qubits with $N>>1$ and will divide this into subsystem A and subsystem B which will each be comprised of N qubits and can be in mixed states. I will also restrict the full AB system to always being in a pure state.

The definitions of complexity that I will utilize depend on a universal set of elementary gates which can approximate unitary operations on our system. The arguments of this paper will not depend on which universal set is used although the actual values for the complexities slightly depend on this choice. The definition of operator complexity as defined in \cite{Nielsen:2006:GAQ:2011686.2011688} is the minimum number of elementary gates required to approximate a unitary operator. From this, the complexity of a target quantum state can be defined: state complexity is the minimal operator complexity of any unitary operator that takes the system from some reference state (typically the completely unentangled state, $\ket{0}^{\otimes 2N}$) to the target state.  

Note that operator complexity and state complexity, although related are not exactly the same. In particular, a unitary operator that will transform the reference state to some target state is not always the most efficient one. The maximum possible complexity\footnote{I have dropped the dependence on $\epsilon$, an error tolerance needed in approximating the unitary transformations in order to keep the complexities finite. I will drop the $\epsilon$ dependence throughout this paper.} of a pure state of N qubits is $2^N$ \cite{Susskind:2015toa}, while the maximum possible complexity of a $2^N$ by $2^N$ unitary operation is $2^{2N}$. This distinction between operator and state complexity will be important when I begin considering density matrices. 

This definition for state complexity is only capable of describing pure states. One of the leading candidates for subsystem complexity proposed in \cite{Agon:2018zso} is what the authors call purification complexity. Purification complexity for a mixed state is defined as the pure state complexity of a purified state minimizing over all possible purifications \footnote{With minimal ancilla qubits used.}. 
The authors showed that this definition possibly has a good dual holographic description.
In the following work, $C(\rho)$ will refer to the purification complexity for a density matrix. The unitary operations that are used to define, $C(\rho_A)$ and $C(\rho_B)$ are done so in our pure $AB$ system.

A notable difference that subsystem complexity has with entropy, is that it does not obey a subadditive relationship. For entropy, subadditivity is:
\begin{equation}
S(\rho_{AB}) \leq S(\rho_A) + S(\rho_B)
\end{equation}
A simple example of complexity not obeying subadditivity is a maximally complex state in the full AB system and reduces to a maximally mixed state when tracing out either subsystem A or B. The maximally complex state has complexity $2^2N$ but each subsystem would only have complexity equal to N. This example is explained in a bit more detail in section \ref{sec:degeneracy}. 

When thinking about the analog of thermodynamics for complexity, the more important quantity to consider may be uncomplexity, $\Delta C$, which is just defined as the separation of complexity of the state from maximum complexity:
\begin{equation}
\Delta C = C_{max} - C 
\end{equation}
The term, $C_{max}$ is just $2^{N}$ for a pure state of N qubits. It may not be obvious what to use for mixed states. The definition I will use for ${C_{max}}(\rho_A)$ will come from finding the density matrix, $\rho_{A,max}$ with the same eigenspectrum as $\rho_A$ such that ${C}(\rho_{A,max})$ is maximized\footnote{Other obvious choices lead to superadditivity being violated in most cases. I thank Adam Brown for suggesting this choice.}. The justification for defining maximum subsystem complexity in this way\footnote{\cite{Zhao:2017isy} approaches defining mixed state uncomplexity differently. It isn't obvious that their definition will always coincide with what is used here but both capture a similar spirit, keeping the eigenspectrum fixed.}, is that subsystem uncomplexity should be a measure of how complicated remaining operations can be while limiting yourself to acting on one system. Unitary operations acting on one subsystem cannot change its density matrix eigenspectrum.

If uncomplexity could be understood as a resource to perform useful computations then a good definition of subsystem/mixed state uncomplexity should obey a superadditive relation: 
\begin{equation}
\Delta C(\rho_{AB}) \geq \Delta C(\rho_A) + \Delta C(\rho_B)
\end{equation}
If there were situations where this statement were not true, then it would suggest the resources available to do useful computations when restricted to only acting on each subsystem separately is greater than the resources available acting on the combined system.
One goal in this paper is to investigate when superadditivity of complexity seems to hold. I will restrict to cases where system AB is pure although ideally, the relation would hold more generally. 

I will also reference basis complexity, $C_B$ and spectrum complexity, $C_S$ which are also defined in \cite{Agon:2018zso}. These quantities are related to purification complexity as follows:
\begin{equation}
C(\rho) \leq C_S(\rho) + C_B(\rho)
\end{equation}
$C(\rho)$, the minimum number of gates required to prepare $\rho$ for any purification must be bound from above by the number of gates required to prepare a density matrix with the same eigenspectrum, $C_S(\rho)$ added to the number of gates required to prepare the density matrix in the correct basis, $C_B(\rho)$.

\begin{figure}[b]
	\includegraphics[width = 3.4in,trim={1.3cm 17.6cm 5.2cm 3.5cm}]{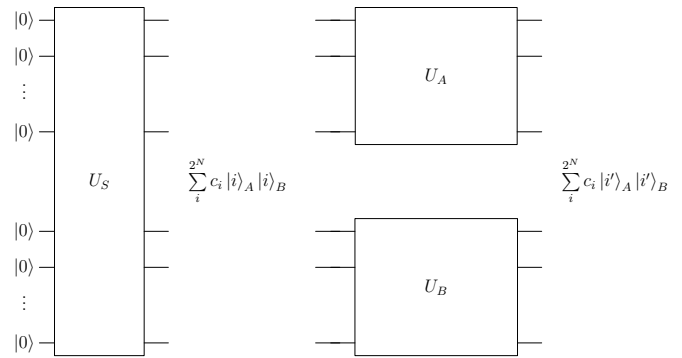}
	\caption{\label{fig:genstate} Any state $\psi_{AB}$ can be prepared with this circuit. Although this may generally be far from the most efficient circuit, the fact it can always be done places an upper bound on the complexity.} 
\end{figure}

I will sometimes focus on basis complexity, $C_B$, by not using the standard reference state (all qubits unentangled and in the 0 state). To explain the prescription for choosing the reference state, first consider the target state, $\ket{\psi_t}$ written in the Schmidt basis:
\begin{equation}
\ket{\psi_t} = \sum_i c_i \ket{i_A}\ket{i_B}
\end{equation}
The reduced density matrix for subsystem A, $\rho_A$ will have eigenvalues, $\lambda_i = |c_i|^2 $ with the same being true for subsystem B. Our non-standard reference state will depend on the coefficients in the Schmidt decomposition, $c_i$. To be specific, our chosen reference state, $\ket{\psi_r}$ will be:
\begin{equation}
\ket{\psi_r} = \sum c_i \ket{i'_A}\ket{i'_B}
\end{equation}
with orthonormal bases $\ket{i'_A}$ and $\ket{i'_B}$ chosen to minimize the reference state's complexity. To reiterate, $\ket{\psi_r}$ is the state\footnote{The complexity of these reference states is similar to what \cite{Agon:2018zso} refers to as the spectrum complexity. Note that it is not exactly the same since I am considering not just a state with the same eigenspectrum, $\lambda_i$ but the same Schmidt coefficients, $c_i$ which has additional phase information. } that has minimal complexity relative to the standard reference state, $\ket{0}^{\otimes2N}$ for any state with Schmidt coefficients, $c_i$. I'll use $\Lambda_A$ or $\Lambda_B$ to denote the reduced density matrices from taking the partial trace of the reference state. Complexity (and uncomplexity) found using this reference state \footnote{From a holographic perspective, an entangled reference state like this might be a natural choice. Such as using the thermofield double state for a two-sided black hole in AdS/CFT.} will always be denoted with a tilde, written as $\tilde{C}$ otherwise the standard reference state will be used. This reference state can be used to describe the complexity of pure states in the full AB system, as well as find the complexities for the subsystems.

This choice of reference state is done to ignore the difficulty of preparing the correct spectrum of eigenvalues. Relative to this reference, finding subsystem complexity, $\tilde{C}(\rho_A)$ is then the minimal number of gates in the minimal unitary transformation, acting only on subsystem A to go from the density matrix, $\Lambda_A$ to $\rho_A$. This is the same as what \cite{Agon:2018zso} calls basis complexity.

There are some important features to note in this definition and properties of the density matrix, $\rho_A$. Every unitary operation on the state, $\Psi_{AB}$ that does not change the eigenspectrum of the reduced states, $\rho_A$ and $\rho_B$ can be decomposed into some $U_A \otimes U_B$. However, these operations $U_A$ and $U_B$ are not always the same as the operators used in finding the complexities of the reduced density matrices $\rho_A$ and $\rho_B$.
As is noted in \cite{Zhao:2017isy}, there will be operations on either subsystem A or B that do not change the density matrix and therefore should not count towards changing its complexity. I will refer to those operations as undetectable by either A or B. In other words, undetectable unitary operations are ones that rotate within the degenerate eigenspace of an eigenvalue and do not change the density matrix. A density matrix's complexity can't be sensitive to undetectable operations.

\section{The Role of Degeneracy} \label{sec:degeneracy}
The degeneracies of density matrices play a surprising role in mixed state complexity. The distinction between detectable and undetectable operations makes maximum subsystem complexity strongly dependent on how degenerate the density matrix is.
First consider constructing a very complex $(\mathcal{O}(2^{2N}))$ state for AB. Any state can be constructed with a circuit as shown in figure \ref{fig:genstate}. This circuit is made up of three unitaries: $U_S$, which prepares the correct eigenspectrum for the density matrices\footnote{I have made the choice to include phase information present in the Schmidt coefficients that the density matrices aren't sensitive to.} and $U_A$ and $U_B$ which transforms to the correct bases for subsystems A and B in the Schmidt decomposition. This circuit is not necessarily the most efficient one to prepare this state, but from the definition of complexity the combined number of elementary gates in $U_S$, $U_A$ and $U_B$ must be greater than or equal to the state's complexity, $(\mathcal{O}(2^{2N}))$.
\begin{equation}
C(U_S) + C(U_A) + C(U_B) \geq \mathcal{O}(2^{2N})
\end{equation}

\begin{figure}[b]
	\includegraphics[width = 3.4in,trim={1.3cm 17.6cm 5.2cm 3.5cm}]{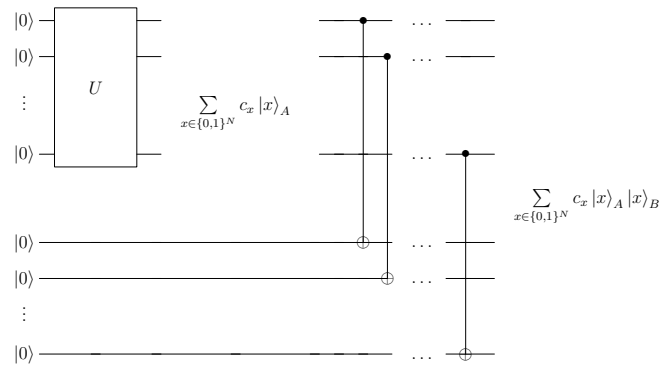}
	\caption{\label{fig:spectrumcircuit} The eigenspectrum of any density matrix, $\rho_A$ can be prepared with this circuit. This particular circuit only prepares density matrices diagonal in the computational basis and its complexity is bounded from above by $2^N + N$ gates.} 
\end{figure}

The spectrum complexity and the complexity of $U_S$ in figure \ref{fig:genstate} cannot be greater than $\mathcal{O}(2^N)$. To see this, consider the circuit in figure \ref{fig:spectrumcircuit} that can be used to construct any density matrix eigenspectrum. This circuit first prepares the desired Schmidt components of the final state in the AB system as the coefficients in the computational basis for a pure state in just the A subsystem. This operation is at most of complexity $\mathcal{O}(2^N)$. Then using N CNOT gates, the desired eigenspectrum is achieved. 
This must mean that the combined $U_A$ and $U_B$ transformations must have had complexity $(\mathcal{O}(2^{2N}))$. The complexities of these operations restricted to acting only on a single subsystem can be very complex, but will not always be relevant for the subsystem's complexity depending on degeneracy.

Very complex states for system AB are expected to reduce to very mixed density matrices for A and B \cite{Page:1993df}. However, as long as the subsystem entropy is nearly maximal, I don't have reason to expect breaking exact degeneracy in the norm of the Schmidt coefficients prevents the AB state from having complexity $\mathcal{O}(2^{2N})$. The degeneracy is relevant to the subsystem complexity because of the distinction between detectable and undetectable operations, but these operators don't have this distinction for the full AB system.

For highly complex AB states, if the density matrices for A and B have no degeneracy, then there are no degenerate subspaces and no undetectable operations. The transformations $U_A$ and $U_B$ will be the minimally complex transformations to prepare the proper basis and the basis complexity of at least one of the density matrices will be $(\mathcal{O}(2^{2N}))$.
Compare that to the maximally mixed state which is very easy to construct in the full AB system. Preparing N bell pairs and assigning each half of the Bell pair to either subsystem A or B, leads to a maximally entangled state. Each Bell pair is prepared with a single Hadamard gate and a single CNOT gate, making the complexity of this full operation, 2N. 
It seems surprising that there can be such a large difference in the subsystem complexity ($\mathcal{O}(2^{2N})$ compared to $\mathcal{O}(N)$) based on breaking exact degeneracy when the full system's pure state complexity wouldn't have to be much different.

\section{Superadditivity} \label{sec:superadditivity}
In this section, I will investigate if the following superadditivity relation holds:
\begin{equation}
\Delta C(\Psi_{AB}) \geq \Delta C(\rho_A) + \Delta C(\rho_B)
\end{equation}

\subsection{Case 1: A and B unentangled:} 
Superadditivity easily holds in this case because keeping subsystems A and B unentangled and pure, severely limits the complexity of the overall state in AB. The complexity of such a state can be at most the sum of the maximal complexities of states in A and states in B (each $2^N$). Maximal complexity grows multiplicatively, not additively, so the maximal complexity of the total system (AB) will always be far larger. $C_{max,AB} - C_{AB} >> C_{max,A} + C_{max,B}$.

\subsection{Case 1A: Many of the eigenvalues are zero:} 
This is similar to case 1, where A and B are unentangled. In this case, the complexity of the full AB state is limited by the average basis complexities of A and B. Let $n_S$ be the Schmidt number (the number of non-zero eigenvalues) of the density matrix, $\rho_A$. The Schmidt decomposition of the state will be: 
\begin{equation}
\Psi_{AB} = \sum \limits_{i=1}^{n_S} c_i\ket{i_A} \otimes \ket{i_B}
\end{equation}
with the terms in the sum being cut off at the Schmidt number.
Making use of a superposition property of complexity which will be described in Section \ref{sec:case3}, the complexity of the AB state is bounded from above:
\begin{eqnarray}
C(\Psi_{AB})\leq \sum \limits_{i=1}^{n_S} C(\ket{i_A}) + C(\ket{i_B}) \leq 2n_S*2^N \\
\Delta C(\Psi_{AB}) \geq 2^{2N} - 2n_S*2^N
\end{eqnarray}
As long as $n_S << 2^N$, this is essentially the same as the unentangled case. The uncomplexities of subsystems A and B will be limited to being $\mathcal{O}(n_S*2^N)$ but the full AB system will have uncomplexity $\mathcal{O}(2^{2N})$. 

\subsection{Case 2: A and B maximally mixed:} 
Unlike the previous cases, these AB states are capable of being maximally complex, $2^{2N}$. As was noted in the previous section, the subsystem complexity for maximally mixed $\rho_A$ and $\rho_B$ is very low, $C(\rho_A) = C(\rho_B) = N$. This might seem problematic since $\Delta C(\Psi_{AB})$ can now be 0. but from the definition of $C_{max}(\rho)$:
\begin{equation}
{\Delta C}(\rho_A) = {\Delta C}(\rho_B) = 0
\end{equation}
This case becomes trivial because $\rho_A$ and $\rho_B$ are proportional to the identity and are therefore diagonal in any basis. Any unitary transformation of the identity matrix will keep it diagonal and are therefore undetectable. The complexity of the density matrix cannot change from a unitary operation only acting on one subsystem (just A or B), so the maximally mixed density matrices are already in their maximally complex state.

Superadditivity is trivially easy to verify in the cases so far, but it provides hints on how to approach the problem more generally. As the density matrix becomes increasingly mixed, the full AB system is capable of being more complex but as the density matrix gains degenerate eigenspaces, the set of detectable operations in only A or only B decreases and therefore the complexity of $\rho_A$ and $\rho_A$ become more limited. The next extreme to consider is a state that is highly entangled state with no eigenspace degeneracy.

\subsection{Case 3: All non-zero eigenvalues and no degeneracy:} \label{sec:case3}
In this situation, the entropy can be arbitrarily close to maximal, but all unitary operations on A are classified as detectable since there are no degenerate eigenspaces. Because of this property, there is a direct connection between the minimal unitary to construct $\ket{\Psi_{AB}}$ and the minimal unitaries to construct $\rho_A$ and $\rho_B$.

\begin{figure}[b]
	\includegraphics[width = 3.4in,trim={1.3cm 17.6cm 5.2cm 3.5cm}]{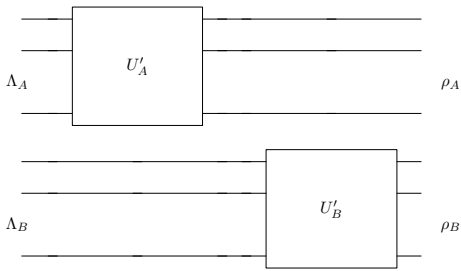}
	\caption{\label{fig:bipartcircuit} In the no-degeneracy case: This is the minimum unitary operations to prepare a state $\ket{\Psi_{AB}}$ starting from the modified reference state. The circuit separates into two disjoint pieces, acting on systems A and B separately.} 
\end{figure}

To show this, I will make use of our modified definition of complexity, $\tilde{C}$. From our modified reference state, call the minimal unitary operations to prepare $\rho_A$ and $\rho_B$, $U_A'$ and $U_B'$ respectively. Normally, there would be many $\ket{\Psi_{AB}}$'s that could reduce to $\rho_A$ and $\rho_B$ after taking the appropriate partial trace. In this case of complete lack of degeneracy however, $\ket{\Psi_{AB}}$ is uniquely picked out (once the phase information in each Schmidt coefficient is included which is contained in the modified reference state). All gates used to prepare $\ket{\Psi_{AB}}$ are detectable by A or B since there are no degenerate subspaces. The minimal unitary transformation to prepare $\ket{\Psi_{AB}}$ is then just $U_A' \otimes U_B'$ since we don't need to prepare the spectrum. This is illustrated in the circuit shown in figure \ref{fig:bipartcircuit}. The complexities of the subsystems and the full system are then additive:
\begin{eqnarray}
\tilde{C}(\rho_A) + \tilde{C}(\rho_B) = \tilde{C}(\Psi_{AB})\\
\tilde{C}(\rho_{A,max}) + \tilde{C}(\rho_{B,max}) = \tilde{C}(\Psi_{AB,max})
\end{eqnarray}
The superadditivity inequality is saturated here using the modified reference state which means it holds for basis complexity. Note that the same may not always be true for purification complexity. 

An interesting point to note is that the distinction between circuit complexity and state complexity disappears when there is no degeneracy. When considering mapping a single pure state to another pure state, there will be a wide range of non-equal unitaries that can accomplish this goal. For example, transforming from the usual reference state, only a single row of the unitary matrix (written in the computational basis) is needed to specify this particular mapping which is why many different unitary operations can accomplish the same task. So when defining pure state complexity, specifically the minimal unitary operation needs to be considered. When transforming density matrices with high non-degeneracy and high Schmidt number however, then how a space of states is being mapped is being specified, not just the transformation of a single state. Starting from the modified reference state, finding the basis complexity of a completely non-degenerate density matrix with all non-zero eigenvalues involves specifying how every basis vector in the Hilbert space is being mapped by the unitary transformation which uniquely determines the unitary transformation.

These are the only cases where I have been able to show superadditivity. I will end with some comments on how to approach the remaining case, when there are some zero eigenvalues and not complete degeneracy:

\subsection{Case 4: Some degeneracy}
The difficulty here that was not present in the non-degenerate case is that the minimal unitary operations to prepare the reduced states $\rho_A$ and $\rho_B$ are not necessarily the same operators used to prepare the full AB state, $\Psi_{AB}$. The degeneracy in $\rho_A$ and $\rho_B$ leads to ambiguity in determining the full state. Because of this, directly comparing complexities $C_{AB}$ and $C_A + C_B$ is difficult. I will motivate a new quantity called average basis complexity as a tool to place bounds on subsystem complexity. 

A useful property of complexity referenced in \cite{Aaronson:2016vto} is how it behaves under superposition \footnote{The reference shows this for the superposition of two orthogonal states, but it is easily extended to more than two states.}. Consider a state, $\psi$ which is in the superposition of some set of orthogonal states, $\{\phi_1,...,\phi_m\}$:
\begin{eqnarray}
\braket{\phi_i|\phi_j} = \delta_{ij} \\
\ket{\psi} = \sum_i c_i \ket{\phi_i}
\end{eqnarray}
then $\psi$'s complexity will be bounded from above by the sum of complexities of those orthogonal states:
\begin{equation}
C(\ket{\psi}) \leq \sum_i C(\phi_i)
\end{equation}

I make use of this property to examine the basis in which the density matrices are diagonalized. Consider an orthonormal basis, $\{ \ket{i} \}$ for the Hilbert space, $\mathcal{H}$. I define the average basis complexity to be the complexity averaged over the pure basis states:
\begin{equation}
C_{avg}(\{ \ket{i}\} ) = \frac{1}{dim(\mathcal{H})}\sum_{i \in \{ \ket{i}\}} C(\ket{i})
\end{equation}

First consider applying this to a pure state in the full AB system written in the Schmidt decomposition:
\begin{equation}
\sum_{i=1}^{n_S}c_i\ket{i_A}\otimes{\ket{i}_B}
\end{equation}
I can place a bound on the complexity of the state in the AB system based on the average basis complexities in the subsystems A and B:
\begin{equation}
C(\ket{\Psi_{AB}}) \leq n_S * (C_{avg}(\{ \ket{i_A}\} )+(C_{avg}(\{ \ket{i_B}\} ))
\end{equation}
therefore, very complex states in the AB system, must have a Schmidt decomposition with high combined average basis complexities for subsystems A and B.

I can also apply this to the density matrices for subsystems A and B. The bases of interest are the bases in which the density matrix is diagonal. If a mixed state in subsystem A is diagonal in the orthonormal basis, ${\ket{i_A}}$, then a possible purification is given by:
\begin{equation}
\sum_{i=1}^{n_S}c_i\ket{i_A}\otimes{\ket{x_i}_B}
\end{equation}
The simple states ${\ket{x_i}_B}$ can all be chosen to have complexity $\mathcal{O}(N)$. Since $\rho_A$'s complexity must be less than this purification, then:
\begin{equation}
C(\rho_A) \leq n_S *(\mathcal{O}(N) + C_{avg}(\{ \ket{i_A}\} )) 
\end{equation}
Unlike the Schmidt decomposition which uniquely picks out bases for A and B, the density matrices for A and B may be diagonal in multiple bases. So the complexity of a density matrix is always limited by its least complex diagonal basis. If the degeneracy of a density matrix allows for a diagonal basis with low average basis complexity, then the complexity of the density matrix can't be large. In the completely degenerate case, there is a diagonalizing basis with the average basis complexity equal to $\mathcal{O}(N)$. In highly degenerate density matrices, bases with most eigenvectors having low complexity should be easy to find.

Next consider the other extreme, a maximally non-degenerate density matrix. From section \ref{sec:degeneracy}, I argued these density matrices can reach $\mathcal{O}(2^{2N})$ complexity. In order to have such high complexities, the basis in which the density matrix is diagonalized must have average basis complexity of $\mathcal{O}(2^N)$. For density matrices with low degeneracy, sometimes only high complexity bases may be possible.

Next I will look at how the average basis complexity of a given Hilbert space can differ for different basis choices\footnote{There is qualitative similarity to \cite{Hashimoto:2018bmb} which also argues for basis dependent complexity. Their reasoning and conclusions however are much different.}. Consider two arbitrary orthonormal bases, $\{ v_i\}$ and $\{ w_i\}$ both spanning, $\mathcal{H}$. Since they span the same space, any vector from one basis can be written as a linear combination of states from the other. Using the superposition property of complexity:
\begin{equation}
C(\ket{w_j}) \leq \sum_{i}C(\ket{v_i}) \ \ \ \ \ \forall j
\end{equation}
Then averaging over all $w_j$'s, gives a relationship between the average basis complexities for the two different bases:
\begin{equation} \label{eq:avgcomp}
C_{avg}( \{ \ket{w_j} \}) \leq dim(\mathcal{H})*C_{avg}(\{ \ket{v_i} \})
\end{equation} 
with the same being true substituting w for v and v for w. If $\{ v_i\}$ is the basis of minimal average basis complexity and $\{ w_i\}$ is the basis of maximal basis complexity for this Hilbert space, then this shows that any basis can have average basis complexity at most $dim(\mathcal{H})$ times larger than any other basis choice.

Applying this property to a density matrix with degenerate eigenspaces with multiplicities, $\{ m_j\}$, further limits can be placed on how much the average basis complexities can vary for two different bases, ($\{ v_{j,i} \}$ and $\{ w_{j,i} \}$) in which the density matrix is diagonal. In this notation,the j index enumerates the eigenspace of the jth eigenvalue and the i indexes basis vectors spanning the $m_j$ dimensional subspace. First I'll write the average basis complexity of the entire Hilbert space, $\mathcal{H}$ for basis, $\{ v_{j,i}\}$ in terms of the average basis complexities of each of the degenerate eigenspaces :
\begin{equation}
C_{avg}(\{ v_{j,i}\}) = \frac{1}{dim(\mathcal{H})}*\sum_{k}m_k * C_{avg}(\{ v_{j=k,i}\})
\end{equation}
In this expression, $\{ v_{j=k,i}\}$ is the set of v basis vectors indexed by i for the kth eigenspace. Next, applying equation \ref{eq:avgcomp} to each eigenspace, the relationship between average basis complexities for the two different bases becomes: 
\begin{equation}
C_{avg}(\{ v_{j,i}\}) \leq \frac{1}{dim(\mathcal{H})}*\sum_{k}m_k^2 * C_{avg}(\{ w_{j=k,i}\}) 
\end{equation} 
This is consistent with what is known about the most extreme cases. With no degeneracy, the basis is fixed and therefore average basis complexity cannot be changed at all. With complete degeneracy, every basis diagonalizes the density matrix and the average basis complexity can range from $\mathcal{O}(N)$ to $2^N$.

Since a density matrix will have its complexity limited by every basis in which it is diagonal, very degenerate density matrices will generally not be able to reach high complexity as long as a low complexity basis exists. For density matrices with low degeneracy, $\mathcal{O}(2^{2N})$ complexity is achievable, and  therefore these matrices must be forced to be diagonalized in a high average complexity basis, $\mathcal{O}(2^{N})$. 

To compare the subsystem complexities to the full system's, consider for example if $\{ \ket{j,i}_A\}$ and $\{ \ket{j,i}_B\}$ are the bases with minimal average basis complexity that $\rho_A$ and $\rho_B$ are diagonal in. Then the Schmidt basis of the full AB system can have average basis complexity of at most, $\frac{1}{n_S}*\sum_{k} m_k^2* (C_{avg}(\{ \ket{j=k,i}_A\}) + C_{avg}(\{ \ket{j=k,i}_B\})$. Applying these as bounds to the complexities gives:
\begin{eqnarray}
C(\rho_A&) &\leq n_S*(\mathcal{O}(N)+ C_{avg}(\{ \ket{j,k}_A\})) \\ 
C(\rho_B&) &\leq n_S*(\mathcal{O}(N)+C_{avg}(\{ \ket{j,k}_B\})) \\
C(\Psi_{AB}&) &\leq  \sum_{k}m_k^2  \\ 
* &(&C_{avg}(\{ \ket{j=k,i}_A\}) + C_{avg}(\{ \ket{j=k,i}_B\}) 
\end{eqnarray} 

To be clear, I have shown how much average basis complexity can differ but have not completely laid out what are the necessary and sufficient conditions for low/high average basis complexities bases to exist. I would like to understand when a low average complexity basis for some subspace isn't possible.

Furthermore, it would be nice to be able to give a more precise statement comparing the complexities of the full system to its subsystems. The average basis complexities restrict the possible values for both system and subsystem complexities. But to really prove superadditivity generally, how $C(\Psi_{AB}) - C(\rho_A) - C(\rho_B)$ varies amongst states of a particular eigenspectrum would need to be known.

\section{Discussion}
In this paper I have investigated properties of a few proposed definitions of subsystem/mixed state complexity: purification and basis complexity. This led to arguments that show the importance of degeneracy of eigenstates of a density matrix for these complexity definitions. The distinction between exact degeneracy and broken degeneracy is surprisingly important and greatly affect how large the complexity of the density matrix can be ($\mathcal{O}(N)$ compared to $\mathcal{O}(2^{2N})$). A density matrix can be thought of as giving a probabilistic combination of pure states. In the case of a maximally degenerate density matrix, there is no inherently preferred basis, which allows for a description made up entirely out of simple (low complexity) states. When little to no degeneracy exists then the density matrix can be forced into a description of a probabilistic mixture of high complexity states. This distinction is closely related to the non-linear nature of complexity for a superposition of states \cite{Maldacena:2013xja, Susskind:2015toa}. A basis of low complexity states can be transformed into a basis of high complexity states and vice-versa with either description (high or low complexity) spanning the same Hilbert space. 

The large separation in complexity scales due to breaking degeneracy reveals a large separation in complexity scales for maximum spectrum and basis complexity. For a $2^N$ dimensional Hilbert space, spectrum complexity must be less than $\mathcal{O}(2^N)$ but basis complexity can reach $\mathcal{O}(2^{2N})$. In other words, states with high uncomplexity have most of the remaining complexity coming from basis complexity.

The full state's complexity has dependence on the average basis complexity in the Schmidt decomposition but there doesn't seem to be reason to believe that the full state's complexity is sensitive to the difference between exact and non-exact degeneracy of it's reduced density matrices. I would like to better understand how a subsystem's complexity is related to the full system's and if it is possible to show superadditivity in complete generality. Understanding this relationship would help in developing a thermodynamic-like description of complexity.

If uncomplexity plays an important role in the formation/existence of black hole firewalls, then understanding subsystem complexity should be required since  composite systems are used to describe black holes as they evaporate or as new things fall past the horizon. When adding a single pure qubit to a large pure system, the maximum possible complexity doubles leading to greatly increased uncomplexity \cite{Brown:2017jil}. While maximum pure state complexity and maximum complexity of very non-degenerate mixed states grow exponentially with number of qubits, highly degenerate density matrices' maximum complexities only grow linearly. An interesting question is if these properties say anything about black hole information or firewalls.

\acknowledgements
I thank Adam Brown for discussions that greatly helped me in learning this topic and motivate understanding subsystem complexity. I also thank Andreas Albrecht, Lenny Susskind, Yasha Neiman and Ying Zhao for helpful discussions. This work was supported by the Quantum Gravity Unit of the Okinawa Institute of Science and
Technology Graduate University (OIST).

\bibliography{complexity}

\begin{thebibliography}{23}%
\makeatletter
\providecommand \@ifxundefined [1]{%
 \@ifx{#1\undefined}
}%
\providecommand \@ifnum [1]{%
 \ifnum #1\expandafter \@firstoftwo
 \else \expandafter \@secondoftwo
 \fi
}%
\providecommand \@ifx [1]{%
 \ifx #1\expandafter \@firstoftwo
 \else \expandafter \@secondoftwo
 \fi
}%
\providecommand \natexlab [1]{#1}%
\providecommand \enquote  [1]{``#1''}%
\providecommand \bibnamefont  [1]{#1}%
\providecommand \bibfnamefont [1]{#1}%
\providecommand \citenamefont [1]{#1}%
\providecommand \href@noop [0]{\@secondoftwo}%
\providecommand \href [0]{\begingroup \@sanitize@url \@href}%
\providecommand \@href[1]{\@@startlink{#1}\@@href}%
\providecommand \@@href[1]{\endgroup#1\@@endlink}%
\providecommand \@sanitize@url [0]{\catcode `\\12\catcode `\$12\catcode
  `\&12\catcode `\#12\catcode `\^12\catcode `\_12\catcode `\%12\relax}%
\providecommand \@@startlink[1]{}%
\providecommand \@@endlink[0]{}%
\providecommand \url  [0]{\begingroup\@sanitize@url \@url }%
\providecommand \@url [1]{\endgroup\@href {#1}{\urlprefix }}%
\providecommand \urlprefix  [0]{URL }%
\providecommand \Eprint [0]{\href }%
\providecommand \doibase [0]{http://dx.doi.org/}%
\providecommand \selectlanguage [0]{\@gobble}%
\providecommand \bibinfo  [0]{\@secondoftwo}%
\providecommand \bibfield  [0]{\@secondoftwo}%
\providecommand \translation [1]{[#1]}%
\providecommand \BibitemOpen [0]{}%
\providecommand \bibitemStop [0]{}%
\providecommand \bibitemNoStop [0]{.\EOS\space}%
\providecommand \EOS [0]{\spacefactor3000\relax}%
\providecommand \BibitemShut  [1]{\csname bibitem#1\endcsname}%
\let\auto@bib@innerbib\@empty
\bibitem [{\citenamefont {Hayden}\ and\ \citenamefont
  {Preskill}(2007)}]{Hayden:2007cs}%
  \BibitemOpen
  \bibfield  {author} {\bibinfo {author} {\bibfnamefont {P.}~\bibnamefont
  {Hayden}}\ and\ \bibinfo {author} {\bibfnamefont {J.}~\bibnamefont
  {Preskill}},\ }\href {\doibase 10.1088/1126-6708/2007/09/120} {\bibfield
  {journal} {\bibinfo  {journal} {JHEP}\ }\textbf {\bibinfo {volume} {09}},\
  \bibinfo {pages} {120} (\bibinfo {year} {2007})},\ \Eprint
  {http://arxiv.org/abs/0708.4025} {arXiv:0708.4025 [hep-th]} \BibitemShut
  {NoStop}%
\bibitem [{\citenamefont {Almheiri}\ \emph {et~al.}(2013)\citenamefont
  {Almheiri}, \citenamefont {Marolf}, \citenamefont {Polchinski},\ and\
  \citenamefont {Sully}}]{Almheiri:2012rt}%
  \BibitemOpen
  \bibfield  {author} {\bibinfo {author} {\bibfnamefont {A.}~\bibnamefont
  {Almheiri}}, \bibinfo {author} {\bibfnamefont {D.}~\bibnamefont {Marolf}},
  \bibinfo {author} {\bibfnamefont {J.}~\bibnamefont {Polchinski}}, \ and\
  \bibinfo {author} {\bibfnamefont {J.}~\bibnamefont {Sully}},\ }\href
  {\doibase 10.1007/JHEP02(2013)062} {\bibfield  {journal} {\bibinfo  {journal}
  {JHEP}\ }\textbf {\bibinfo {volume} {02}},\ \bibinfo {pages} {062} (\bibinfo
  {year} {2013})},\ \Eprint {http://arxiv.org/abs/1207.3123} {arXiv:1207.3123
  [hep-th]} \BibitemShut {NoStop}%
\bibitem [{\citenamefont {Ryu}\ and\ \citenamefont
  {Takayanagi}(2006)}]{Ryu:2006bv}%
  \BibitemOpen
  \bibfield  {author} {\bibinfo {author} {\bibfnamefont {S.}~\bibnamefont
  {Ryu}}\ and\ \bibinfo {author} {\bibfnamefont {T.}~\bibnamefont
  {Takayanagi}},\ }\href {\doibase 10.1103/PhysRevLett.96.181602} {\bibfield
  {journal} {\bibinfo  {journal} {Phys. Rev. Lett.}\ }\textbf {\bibinfo
  {volume} {96}},\ \bibinfo {pages} {181602} (\bibinfo {year} {2006})},\
  \Eprint {http://arxiv.org/abs/hep-th/0603001} {arXiv:hep-th/0603001 [hep-th]}
  \BibitemShut {NoStop}%
\bibitem [{\citenamefont {Faulkner}\ \emph {et~al.}(2014)\citenamefont
  {Faulkner}, \citenamefont {Guica}, \citenamefont {Hartman}, \citenamefont
  {Myers},\ and\ \citenamefont {Van~Raamsdonk}}]{Faulkner:2013ica}%
  \BibitemOpen
  \bibfield  {author} {\bibinfo {author} {\bibfnamefont {T.}~\bibnamefont
  {Faulkner}}, \bibinfo {author} {\bibfnamefont {M.}~\bibnamefont {Guica}},
  \bibinfo {author} {\bibfnamefont {T.}~\bibnamefont {Hartman}}, \bibinfo
  {author} {\bibfnamefont {R.~C.}\ \bibnamefont {Myers}}, \ and\ \bibinfo
  {author} {\bibfnamefont {M.}~\bibnamefont {Van~Raamsdonk}},\ }\href {\doibase
  10.1007/JHEP03(2014)051} {\bibfield  {journal} {\bibinfo  {journal} {JHEP}\
  }\textbf {\bibinfo {volume} {03}},\ \bibinfo {pages} {051} (\bibinfo {year}
  {2014})},\ \Eprint {http://arxiv.org/abs/1312.7856} {arXiv:1312.7856
  [hep-th]} \BibitemShut {NoStop}%
\bibitem [{\citenamefont {Swingle}\ and\ \citenamefont
  {Van~Raamsdonk}(2014)}]{Swingle:2014uza}%
  \BibitemOpen
  \bibfield  {author} {\bibinfo {author} {\bibfnamefont {B.}~\bibnamefont
  {Swingle}}\ and\ \bibinfo {author} {\bibfnamefont {M.}~\bibnamefont
  {Van~Raamsdonk}},\ }\href@noop {} {\  (\bibinfo {year} {2014})},\ \Eprint
  {http://arxiv.org/abs/1405.2933} {arXiv:1405.2933 [hep-th]} \BibitemShut
  {NoStop}%
\bibitem [{\citenamefont {Stanford}\ and\ \citenamefont
  {Susskind}(2014)}]{Stanford:2014jda}%
  \BibitemOpen
  \bibfield  {author} {\bibinfo {author} {\bibfnamefont {D.}~\bibnamefont
  {Stanford}}\ and\ \bibinfo {author} {\bibfnamefont {L.}~\bibnamefont
  {Susskind}},\ }\href {\doibase 10.1103/PhysRevD.90.126007} {\bibfield
  {journal} {\bibinfo  {journal} {Phys. Rev.}\ }\textbf {\bibinfo {volume}
  {D90}},\ \bibinfo {pages} {126007} (\bibinfo {year} {2014})},\ \Eprint
  {http://arxiv.org/abs/1406.2678} {arXiv:1406.2678 [hep-th]} \BibitemShut
  {NoStop}%
\bibitem [{\citenamefont {Brown}\ \emph {et~al.}(2016)\citenamefont {Brown},
  \citenamefont {Roberts}, \citenamefont {Susskind}, \citenamefont {Swingle},\
  and\ \citenamefont {Zhao}}]{Brown:2015bva}%
  \BibitemOpen
  \bibfield  {author} {\bibinfo {author} {\bibfnamefont {A.~R.}\ \bibnamefont
  {Brown}}, \bibinfo {author} {\bibfnamefont {D.~A.}\ \bibnamefont {Roberts}},
  \bibinfo {author} {\bibfnamefont {L.}~\bibnamefont {Susskind}}, \bibinfo
  {author} {\bibfnamefont {B.}~\bibnamefont {Swingle}}, \ and\ \bibinfo
  {author} {\bibfnamefont {Y.}~\bibnamefont {Zhao}},\ }\href {\doibase
  10.1103/PhysRevLett.116.191301} {\bibfield  {journal} {\bibinfo  {journal}
  {Phys. Rev. Lett.}\ }\textbf {\bibinfo {volume} {116}},\ \bibinfo {pages}
  {191301} (\bibinfo {year} {2016})},\ \Eprint
  {http://arxiv.org/abs/1509.07876} {arXiv:1509.07876 [hep-th]} \BibitemShut
  {NoStop}%
\bibitem [{\citenamefont {Carmi}\ \emph {et~al.}(2017)\citenamefont {Carmi},
  \citenamefont {Chapman}, \citenamefont {Marrochio}, \citenamefont {Myers},\
  and\ \citenamefont {Sugishita}}]{Carmi:2017jqz}%
  \BibitemOpen
  \bibfield  {author} {\bibinfo {author} {\bibfnamefont {D.}~\bibnamefont
  {Carmi}}, \bibinfo {author} {\bibfnamefont {S.}~\bibnamefont {Chapman}},
  \bibinfo {author} {\bibfnamefont {H.}~\bibnamefont {Marrochio}}, \bibinfo
  {author} {\bibfnamefont {R.~C.}\ \bibnamefont {Myers}}, \ and\ \bibinfo
  {author} {\bibfnamefont {S.}~\bibnamefont {Sugishita}},\ }\href {\doibase
  10.1007/JHEP11(2017)188} {\bibfield  {journal} {\bibinfo  {journal} {JHEP}\
  }\textbf {\bibinfo {volume} {11}},\ \bibinfo {pages} {188} (\bibinfo {year}
  {2017})},\ \Eprint {http://arxiv.org/abs/1709.10184} {arXiv:1709.10184
  [hep-th]} \BibitemShut {NoStop}%
\bibitem [{\citenamefont {Harlow}\ and\ \citenamefont
  {Hayden}(2013)}]{Harlow:2013tf}%
  \BibitemOpen
  \bibfield  {author} {\bibinfo {author} {\bibfnamefont {D.}~\bibnamefont
  {Harlow}}\ and\ \bibinfo {author} {\bibfnamefont {P.}~\bibnamefont
  {Hayden}},\ }\href {\doibase 10.1007/JHEP06(2013)085} {\bibfield  {journal}
  {\bibinfo  {journal} {JHEP}\ }\textbf {\bibinfo {volume} {06}},\ \bibinfo
  {pages} {085} (\bibinfo {year} {2013})},\ \Eprint
  {http://arxiv.org/abs/1301.4504} {arXiv:1301.4504 [hep-th]} \BibitemShut
  {NoStop}%
\bibitem [{\citenamefont {Maldacena}\ and\ \citenamefont
  {Susskind}(2013)}]{Maldacena:2013xja}%
  \BibitemOpen
  \bibfield  {author} {\bibinfo {author} {\bibfnamefont {J.}~\bibnamefont
  {Maldacena}}\ and\ \bibinfo {author} {\bibfnamefont {L.}~\bibnamefont
  {Susskind}},\ }\href {\doibase 10.1002/prop.201300020} {\bibfield  {journal}
  {\bibinfo  {journal} {Fortsch. Phys.}\ }\textbf {\bibinfo {volume} {61}},\
  \bibinfo {pages} {781} (\bibinfo {year} {2013})},\ \Eprint
  {http://arxiv.org/abs/1306.0533} {arXiv:1306.0533 [hep-th]} \BibitemShut
  {NoStop}%
\bibitem [{\citenamefont {Brown}\ and\ \citenamefont
  {Susskind}(2018)}]{Brown:2017jil}%
  \BibitemOpen
  \bibfield  {author} {\bibinfo {author} {\bibfnamefont {A.~R.}\ \bibnamefont
  {Brown}}\ and\ \bibinfo {author} {\bibfnamefont {L.}~\bibnamefont
  {Susskind}},\ }\href {\doibase 10.1103/PhysRevD.97.086015} {\bibfield
  {journal} {\bibinfo  {journal} {Phys. Rev.}\ }\textbf {\bibinfo {volume}
  {D97}},\ \bibinfo {pages} {086015} (\bibinfo {year} {2018})},\ \Eprint
  {http://arxiv.org/abs/1701.01107} {arXiv:1701.01107 [hep-th]} \BibitemShut
  {NoStop}%
\bibitem [{\citenamefont {Gour}\ \emph {et~al.}(2013)\citenamefont {Gour},
  \citenamefont {P.~Müller}, \citenamefont {Narasimhachar}, \citenamefont
  {Spekkens},\ and\ \citenamefont {Yunger~Halpern}}]{Gour:2013}%
  \BibitemOpen
  \bibfield  {author} {\bibinfo {author} {\bibfnamefont {G.}~\bibnamefont
  {Gour}}, \bibinfo {author} {\bibfnamefont {M.}~\bibnamefont {P.~Müller}},
  \bibinfo {author} {\bibfnamefont {V.}~\bibnamefont {Narasimhachar}}, \bibinfo
  {author} {\bibfnamefont {R.}~\bibnamefont {Spekkens}}, \ and\ \bibinfo
  {author} {\bibfnamefont {N.}~\bibnamefont {Yunger~Halpern}},\ }\bibfield
  {booktitle} {\emph {\bibinfo {booktitle} {Physics Reports}},\ }\href@noop {}
  {\ \textbf {\bibinfo {volume} {583}} (\bibinfo {year} {2013})}\BibitemShut
  {NoStop}%
\bibitem [{\citenamefont {Alishahiha}(2015)}]{Alishahiha:2015rta}%
  \BibitemOpen
  \bibfield  {author} {\bibinfo {author} {\bibfnamefont {M.}~\bibnamefont
  {Alishahiha}},\ }\href {\doibase 10.1103/PhysRevD.92.126009} {\bibfield
  {journal} {\bibinfo  {journal} {Phys. Rev.}\ }\textbf {\bibinfo {volume}
  {D92}},\ \bibinfo {pages} {126009} (\bibinfo {year} {2015})},\ \Eprint
  {http://arxiv.org/abs/1509.06614} {arXiv:1509.06614 [hep-th]} \BibitemShut
  {NoStop}%
\bibitem [{\citenamefont {Ben-Ami}\ and\ \citenamefont
  {Carmi}(2016)}]{Ben-Ami:2016qex}%
  \BibitemOpen
  \bibfield  {author} {\bibinfo {author} {\bibfnamefont {O.}~\bibnamefont
  {Ben-Ami}}\ and\ \bibinfo {author} {\bibfnamefont {D.}~\bibnamefont
  {Carmi}},\ }\href {\doibase 10.1007/JHEP11(2016)129} {\bibfield  {journal}
  {\bibinfo  {journal} {JHEP}\ }\textbf {\bibinfo {volume} {11}},\ \bibinfo
  {pages} {129} (\bibinfo {year} {2016})},\ \Eprint
  {http://arxiv.org/abs/1609.02514} {arXiv:1609.02514 [hep-th]} \BibitemShut
  {NoStop}%
\bibitem [{\citenamefont {Agon}\ \emph {et~al.}(2018)\citenamefont {Agon},
  \citenamefont {Headrick},\ and\ \citenamefont {Swingle}}]{Agon:2018zso}%
  \BibitemOpen
  \bibfield  {author} {\bibinfo {author} {\bibfnamefont {C.~A.}\ \bibnamefont
  {Agon}}, \bibinfo {author} {\bibfnamefont {M.}~\bibnamefont {Headrick}}, \
  and\ \bibinfo {author} {\bibfnamefont {B.}~\bibnamefont {Swingle}},\
  }\href@noop {} {\  (\bibinfo {year} {2018})},\ \Eprint
  {http://arxiv.org/abs/1804.01561} {arXiv:1804.01561 [hep-th]} \BibitemShut
  {NoStop}%
\bibitem [{\citenamefont {Jefferson}\ and\ \citenamefont
  {Myers}(2017)}]{Jefferson:2017sdb}%
  \BibitemOpen
  \bibfield  {author} {\bibinfo {author} {\bibfnamefont {R.}~\bibnamefont
  {Jefferson}}\ and\ \bibinfo {author} {\bibfnamefont {R.~C.}\ \bibnamefont
  {Myers}},\ }\href {\doibase 10.1007/JHEP10(2017)107} {\bibfield  {journal}
  {\bibinfo  {journal} {JHEP}\ }\textbf {\bibinfo {volume} {10}},\ \bibinfo
  {pages} {107} (\bibinfo {year} {2017})},\ \Eprint
  {http://arxiv.org/abs/1707.08570} {arXiv:1707.08570 [hep-th]} \BibitemShut
  {NoStop}%
\bibitem [{\citenamefont {Chapman}\ \emph {et~al.}(2018)\citenamefont
  {Chapman}, \citenamefont {Heller}, \citenamefont {Marrochio},\ and\
  \citenamefont {Pastawski}}]{Chapman:2017rqy}%
  \BibitemOpen
  \bibfield  {author} {\bibinfo {author} {\bibfnamefont {S.}~\bibnamefont
  {Chapman}}, \bibinfo {author} {\bibfnamefont {M.~P.}\ \bibnamefont {Heller}},
  \bibinfo {author} {\bibfnamefont {H.}~\bibnamefont {Marrochio}}, \ and\
  \bibinfo {author} {\bibfnamefont {F.}~\bibnamefont {Pastawski}},\ }\href
  {\doibase 10.1103/PhysRevLett.120.121602} {\bibfield  {journal} {\bibinfo
  {journal} {Phys. Rev. Lett.}\ }\textbf {\bibinfo {volume} {120}},\ \bibinfo
  {pages} {121602} (\bibinfo {year} {2018})},\ \Eprint
  {http://arxiv.org/abs/1707.08582} {arXiv:1707.08582 [hep-th]} \BibitemShut
  {NoStop}%
\bibitem [{\citenamefont {Nielsen}(2006)}]{Nielsen:2006:GAQ:2011686.2011688}%
  \BibitemOpen
  \bibfield  {author} {\bibinfo {author} {\bibfnamefont {M.~A.}\ \bibnamefont
  {Nielsen}},\ }\href {http://dl.acm.org/citation.cfm?id=2011686.2011688}
  {\bibfield  {journal} {\bibinfo  {journal} {Quantum Info. Comput.}\ }\textbf
  {\bibinfo {volume} {6}},\ \bibinfo {pages} {213} (\bibinfo {year}
  {2006})}\BibitemShut {NoStop}%
\bibitem [{\citenamefont {Susskind}(2016)}]{Susskind:2015toa}%
  \BibitemOpen
  \bibfield  {author} {\bibinfo {author} {\bibfnamefont {L.}~\bibnamefont
  {Susskind}},\ }\href {\doibase 10.1002/prop.201500091} {\bibfield  {journal}
  {\bibinfo  {journal} {Fortsch. Phys.}\ }\textbf {\bibinfo {volume} {64}},\
  \bibinfo {pages} {84} (\bibinfo {year} {2016})},\ \Eprint
  {http://arxiv.org/abs/1507.02287} {arXiv:1507.02287 [hep-th]} \BibitemShut
  {NoStop}%
\bibitem [{\citenamefont {Zhao}(2018)}]{Zhao:2017isy}%
  \BibitemOpen
  \bibfield  {author} {\bibinfo {author} {\bibfnamefont {Y.}~\bibnamefont
  {Zhao}},\ }\href {\doibase 10.1103/PhysRevD.97.126007} {\bibfield  {journal}
  {\bibinfo  {journal} {Phys. Rev.}\ }\textbf {\bibinfo {volume} {D97}},\
  \bibinfo {pages} {126007} (\bibinfo {year} {2018})},\ \Eprint
  {http://arxiv.org/abs/1711.03125} {arXiv:1711.03125 [hep-th]} \BibitemShut
  {NoStop}%
\bibitem [{\citenamefont {Page}(1993)}]{Page:1993df}%
  \BibitemOpen
  \bibfield  {author} {\bibinfo {author} {\bibfnamefont {D.~N.}\ \bibnamefont
  {Page}},\ }\href {\doibase 10.1103/PhysRevLett.71.1291} {\bibfield  {journal}
  {\bibinfo  {journal} {Phys. Rev. Lett.}\ }\textbf {\bibinfo {volume} {71}},\
  \bibinfo {pages} {1291} (\bibinfo {year} {1993})},\ \Eprint
  {http://arxiv.org/abs/gr-qc/9305007} {arXiv:gr-qc/9305007 [gr-qc]}
  \BibitemShut {NoStop}%
\bibitem [{\citenamefont {Aaronson}(2016)}]{Aaronson:2016vto}%
  \BibitemOpen
  \bibfield  {author} {\bibinfo {author} {\bibfnamefont {S.}~\bibnamefont
  {Aaronson}}\ }(\bibinfo {year} {2016})\ \Eprint
  {http://arxiv.org/abs/1607.05256} {arXiv:1607.05256 [quant-ph]} \BibitemShut
  {NoStop}%
\bibitem [{\citenamefont {Hashimoto}\ \emph {et~al.}(2018)\citenamefont
  {Hashimoto}, \citenamefont {Iizuka},\ and\ \citenamefont
  {Sugishita}}]{Hashimoto:2018bmb}%
  \BibitemOpen
  \bibfield  {author} {\bibinfo {author} {\bibfnamefont {K.}~\bibnamefont
  {Hashimoto}}, \bibinfo {author} {\bibfnamefont {N.}~\bibnamefont {Iizuka}}, \
  and\ \bibinfo {author} {\bibfnamefont {S.}~\bibnamefont {Sugishita}},\
  }\href@noop {} {\  (\bibinfo {year} {2018})},\ \Eprint
  {http://arxiv.org/abs/1805.04226} {arXiv:1805.04226 [hep-th]} \BibitemShut
  {NoStop}%
\end{thebibliography}%

\end{document}